\documentclass[aps,pre,reprint,showpacs]{revtex4-1}
\usepackage[dvips]{graphicx}
\usepackage{amssymb}

\begin{document}

\title{Intrinsic life-time and external manipulation of N\'{e}el states in antiferromagnetic adatom spins on semiconductor surfaces}
\author{Jun Li}
\author{Bang-Gui Liu}
\email[Corresponding author:~]{bgliu@iphy.ac.cn}
\affiliation{Beijing National Laboratory for Condensed Matter
Physics, Institute of Physics, Chinese Academy of Sciences, Beijing
100190, China}

\date{\today }

\begin{abstract}
It has been proposed that antiferromagnetic Fe adatom spins on
semiconductor Cu-N surfaces can be used to store information [S.
Loth {\it et al}, Science \textbf{335}, 196 (2012)]. Here, we
investigate spin dynamics of such antiferromagnetic systems through
Monte Carlo simulations. We find out the temperature and size laws
of switching rates of N\'{e}el states and show that the N\'{e}el
states can become stable enough for the information storage when the number of
spins reaches to one or two dozens of the Fe spins. We also explore
promising methods for manipulating the N\'{e}el states. These could
help realize information storage with such antiferromagnetic spin
systems.
\end{abstract}

\pacs{75.75.-c, 75.78.-n, 75.10.-b, 75.90.+w}

\maketitle

\section{Introduction}

The atomic-size antiferromagnets (AFM)
on semiconductor surfaces have been realized thanks
to technology advance, and have attracted a lot of attention for
their potential in magnetic storage
technology\cite{ATM1,ATM2,ATM3,ATM4,add1}. Due to insensitivity to
magnetic fields, these nanomagnets are considered to be able to meet
the needs of the vast growth of storage density in magnetic
media\cite {AFM3}. On the other hand, it is also far more difficult
to effectively manipulate the AFM stagger magnetization (or N\'{e}el
state) than ferromagnetic magnetization in ferromagnetic materials.
Modern scanning tunneling microscopy (STM) technology can be used to manipulate magnetic states of such nanomagnets through applying an electric field or injecting an electron
current on one of adatom spins\cite{AFM3,STM1,STM2,AFM4,add1}. It has been reported that AFM
chains assembled by placing Fe adatoms on a Cu$_2$N overlayer on Cu(100)
can be switched between two quasistable N\'{e}el states when the
polarized electrons are made to flow across one atom of the
chain\cite{AFM4}. There have been theoretical studies on
magnetic transitions induced by tunneling electrons\cite{THEORY1,THEORY2,THEORY3,THEORY4,THEORY5}, but
temperature effects need to be considered. A reliable study on the temperature-dependent
dynamics of N\'{e}el states is in need.

Here, we use the dynamic Monte-Carlo method to study the
switching rates of a series of such antiferromagnetic Fe spin
chains. We investigate effects of various temperatures and
spin-polarized currents in terms of experimental parameters.
Arrhenius-like behaviors are observed until the temperature is too
low to hurdle the thermal activation barrier of the spin reversal. Our results show that such spin
chains can be stable enough for practical usage when one or two dozen Fe adatoms are
used. We also explore effective methods for external manipulation of
the N\'{e}el states. More detailed results will be presented in the
following.

\section{Spin Model and simulation method}

For low temperatures ($T < 10$ K), the Fe spin bi-chains on the CuN/Cu(100)
surface can be effectively described with an anisotropic Heisenberg
antiferromagnetic model of $S=2$\cite{AFM4,STM1},
\begin{equation}
\hat{H}_{0}=\sum_{i,j}\left[\hat{H}^{0}_{i,j}-J\hat{\vec{S}}_{i,j}\cdot
\hat{\vec{S}}_{i+1,j}\right] -\sum_{i}J^{\prime} \hat{\vec{S}}_{i,1}\cdot
\hat{\vec{S}}_{i,2}, \label{e:1}
\end{equation}
where
\begin{equation}
 \hat{H}^{0}_{i,j}= -g\mu _{B}\vec{B}\cdot\hat{\vec{S}}_{i,j}
-D(\hat{S}_{i,j}^{z})^{2}-E[(\hat{S}_{i,j}^{x})^{2}-(\hat{S}_{i,j}^{y})^{2}]
\end{equation}
describes the $i$-th Fe spin in the $j$-th chain ($j$ = 1, 2). Here,
$\hat{\vec{S}}_{i,j}=\{\hat{S}_{i,j}^{x},\hat{S}_{i,j}^{y},\hat{S}_{i,j}^{z}\}$
is the spin vector operator for the single Fe spin, and as shown in Fig. 1, $J$ is the nearest
antiferromagnetic exchange constant in one of the chains, and $J^{\prime}$
the ferromagnetic exchange constant between the two chains. Letting $J^\prime=0$ and confining $j$ to 1 in (1), we can obtain a single chain system. The parameter $g$ is the Lande g factor (here $g=2$ is used), and $\mu _{B}$ the Bohr magneton. The
parameters $D$ and $E$ are used to characterize the single-ion
magnetic anisotropy, and $\vec{B}=(B^x,B^y,B^z)$ is the applied
magnetic field. In addition to (1), an effective magnetic field $\vec{B}_{tip}$ needs to be applied on the first spin $\hat{\vec{S}}_{1,1}$ to split the two N\'{e}el states and some spin polarized electron currents must be injected through $\hat{\vec{S}}_{1,1}$ to control and detect the N\'{e}el states, as have been done in the previous experimental and theoretical papers\cite{AFM4,THEORY5}.

\begin{figure}[tbp]
\includegraphics[width=8cm]{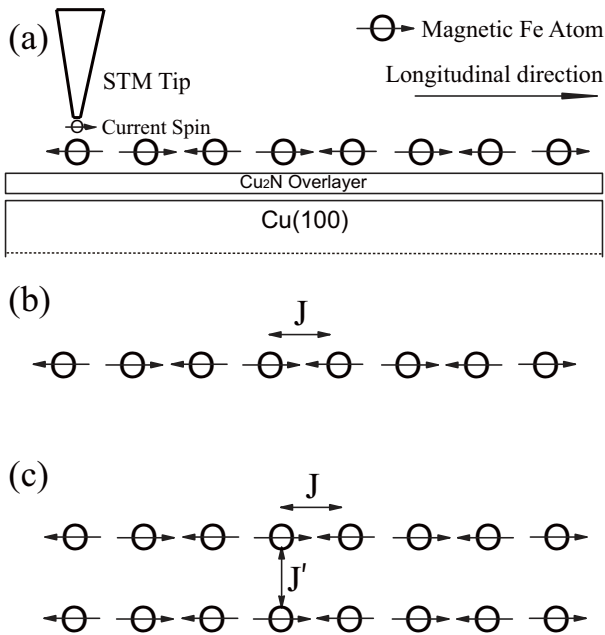}
\caption{(a) A schematic of Fe adatom spins on a Cu$_2$N overlayer
on Cu(100), with electronic current injected through the STM tip.
(b) The inter-spin exchange constant along the chain is $J$. (c) The
inter-spin exchange constant between the two chains is
$J^{\prime}$.} \label{fig:1}
\end{figure}

We use the theoretical methods by Gauyacq {\it et al}\cite{THEORY5} to treat the effects of injected electrons on the spins in the chain and bi-chain systems.
When we treat a spin at the ($i$,$j$) site to calculate the thermal activation barrier, we use mean-field
approximation for its
nearest spins, $\hat{\vec{S}}_{i+1,j}=\vec{S}_{i+1,j}$. This results in effective Hamiltonian
$\hat{H}_0=\sum_{i,j}\hat{H}_{i,j}$, where
$\hat{H}_{i,j}=-\hat{\vec{S}}_{i,j}\cdot
\vec{F}_{i,j}-D(\hat{S}_{i,j}^{z})^{2}-E[(\hat{S}_{i,j}^{x})^{2}-(\hat{S}_{i,j}^{y})^{2}]$
with $\vec{F}_{i,j}$ defined as $J \vec{S}_{i+1,j}+J^{\prime}
\vec{S}_{i,j+1}+g\mu_{B}\vec{B}$. Here, $\vec{S}_{i+1,j}$ and $\vec{S}_{i,j+1}$ are already classical quantities, with the transverse components being equivalent to zero. Their starting values are set in terms of the N\'{e}el states (more detail will be given in the following), and they are updated at each step of the Monte Carlo simulation. For each electron tunneling from the STM tip through the first Fe atom and into the substrate, there is a chance to change the states of the whole Fe spin chains.

For our DMC simulation, the MC steps are defined by the time points:
$t_{n}=\Delta t\cdot n$, where $n$ takes non-negative integers in
sequence. The $\Delta t$ is set to $1.6\times 10^{-7}$ s or determined by $\Delta t= 1e/I$, where $I$ is
the current intensity from the STM tips. $\Delta t$ is chosen to
satisfy the requirement that there is only one electron within a MC
step. At the beginning, we set all of the spins at one for N\'{e}el
state $\acute{E}_{0}$, and the $N$th spin should be set as
$(-1)^{N}\cdot S$. For the $n$th step, the MC simulations are
performed in the following way.

For each Fe spin, we take $\hat{H}_{i,j}$ as a 5 states quantum
model, and there is a thermal-activated energy barrier $\Delta
e$ between the starting and ending states. Using the Arrhenius
law\cite{ArrheniusLaw2,glauber}, we obtain the thermal-activated
spin reversal probability, $P_{t}=1-\exp (r\Delta t)$, for each MC
step\cite{dmc1,dmc2,lbg}, where $r=r_{0}\exp (\frac{-\Delta
e}{k_{B}T})$ is the spin reversal rate and $r_{0}$ is the character
attempt frequency. When an injected electron is tunneling through
the first Fe spin, the collision channel states $S_{c}$ are defined
to simulate the electron-Fe-spin coupling, and the
electron-activated spin reversal probability is defined as
$P_{e}$\cite{THEORY5,ABT1,ETA1,ETA2,ETA3}. Therefore, the total probability for the first Fe spin
is equivalent to $P^1_{t}=1-(1-P_{e})(1-P_{t})$. Actually, at extra-low temperature an additional collective spin channel becomes available for the system to transit from one N\'{e}el state to the other. This transition probability, calculated through exact diagonalization of the Hamiltonian, is added to the above probability to completely describe the spin dynamics.

For the $n$th MC step, each spin in the system has chance to reverse.
With $n$ increasing, at last, the system finally transits to the other N\'{e}el state $\acute{E}_{1}$ at a special $n$.
This $n$ value is denoted by $N_r$. We define
$\tau_1=\langle N_r\rangle \cdot\Delta t$ as the average transition
time from $\acute{E}_{0}$ to $\acute{E}_{1}$. The average transition
time from $\acute{E}_{1}$ to $\acute{E}_{0}$, $\tau_2$, is
calculated in the same way. The switching rate is defined as
$R=2/(\tau_1+\tau_2)$. To be consistent with experiment\cite{AFM4,STM1}, we take
$J$=-0.737 meV, $J^{\prime}$=0.03 meV, $D$=1.87 meV, $E$=0.061 meV, $B^z_{tip}=0.115$ tesla,
and $r_0$=2$\times $10$^{8}s^{-1}$ in the following. Our main
results are calculated by averaging over 100,000 runs of DMC
simulations.

\begin{figure}[tbp]
\includegraphics[width=8cm]{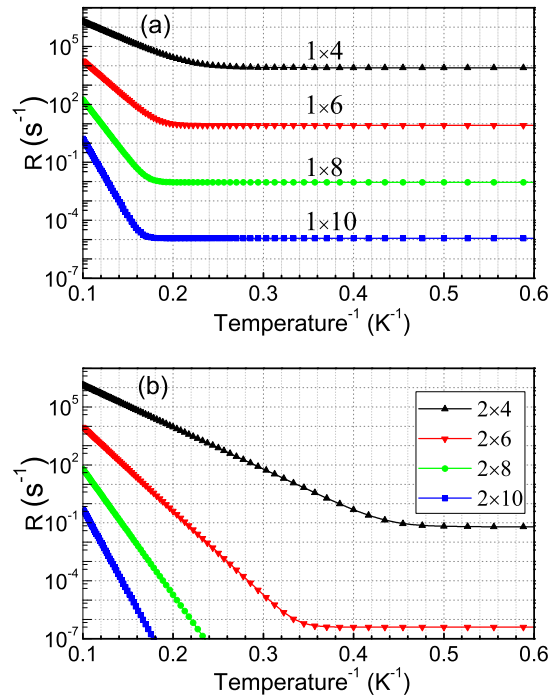}
\caption{Temperature dependence of the switching rates without
magnetic field for $1\times N$ linear chains with $N$ = 4, 6, 8, and
10 (a); and $2\times N$ bi-chains with $N$ = 4, 6, 8, and 10 (b). }
\label{fig:2}
\end{figure}

\section{Intrinsic life times of N\'{e}el states}

We simulate
the spin dynamics for the different Fe chains with
magnetic fields. It is confirmed that the magnetic field has little
effect on switching rates. Simulated switching rates for $1\times N$
Fe-spin chains are presented in Fig. \ref{fig:2}(a). It is clear
that the switching rates show strong temperature dependence as
expected. Above 6 K, the switching rates of the $1\times N$ Fe
chains mainly follow the Arrhenius law. With the addition of two Fe
spins, the energy barrier increases with 4.15 meV. The switching
rates of N\'{e}el states in long Fe chains are tiny in comparison
with those in short ones. When temperature decreases away from the
Arrhenius regime, the switching rates become independent of
temperature. Such behavior can be attributed to the quantum
tunneling because the thermal activity is frozen out at those
ultra-low temperatures. Simulated switching rates of $2\times N$
chains are shown in Fig. \ref{fig:2}(b). In these cases, the
Arrhenius law will work well until the temperature is lower than
about 3 K, and the quantum tunneling rates become so small that the
N\'{e}el states can be considered to be quite stable below about 2
K. It should be pointed that the simulated temperature-rate curves of single chains for $1\times 6$ and $1\times 8$ are in good agreement with experimental curves and those of bi-chains for $2\times 4$ and $2\times 6$ are also consistent with experimental results\cite{AFM4}, which show that our methods and results are both reliable.

It is very interesting that the temperature-dependent switching
rates can be fitted with the simple function:
\begin{equation}\label{eq:R}
R=R_{0}\exp (\frac{-\Delta E}{k_{B}T})+R_{T},
\end{equation}
where $R_0$, $\Delta E$, and $R_T$ are fitting parameters. The
intrinsic life time of the N\'{e}el state can be described with
$\tau=1/R$. For $T=0$, we get $\tau=\tau_0=1/R_T$, the low
temperature limit of the life time. For the $1\times N$ chains,
since $R_0$ is nearly constant, $R_0\sim 2.0\times 10^{-8}$ per
second, we present $\Delta E$, $R_T$, and $\tau$ (for T=0, 8, 10K)
as functions of chain length in Fig. 3. Simple fitting leads to the
following important results.
\begin{equation}
 \Delta E=2.07\cdot N-4.37, ~~~ \tau_0=2.97\times 10^{-10}\times
 27.9^N,
\end{equation}
where $\Delta E$ is in meV and $\tau_0$ in second (s). The tunneling
rate $R_T$ will reduce by 778-fold when two Fe adatoms are added, which is in good agreement with experimental value\cite{AFM4}.
With all the three parameters known, we can calculate life times
($\tau=1/R$) of N\'{e}el states for any given temperature and chain
length. For the ultra-low temperature limit (0K), we have 0.12 s,
1.8 minute, 1.0 day, and 2.1 year for the $1\times N$ chains with
$N$=6, 8, 10, and 12, respectively. For the $2\times N$ chains, we
have $R_0=2.0\sim 2.2\times 10^{-8}$ per second, and the $N$
dependence relations of $\Delta E$ and $\tau_0$ ($R_T$) are similar
to Equ. (4). For a special case of $2\times N$ ($N=8$) chains, we already
have $\tau_0$ $\sim$ 28.2 day. Although increasing temperature will
decrease the life time, very long life time $\tau$ of a few years
can be achieved as long as $N$ is large enough, making $1\times 14$
and $2\times 8$ or longer chains.

\begin{figure}[tbp]
\includegraphics[width=8cm]{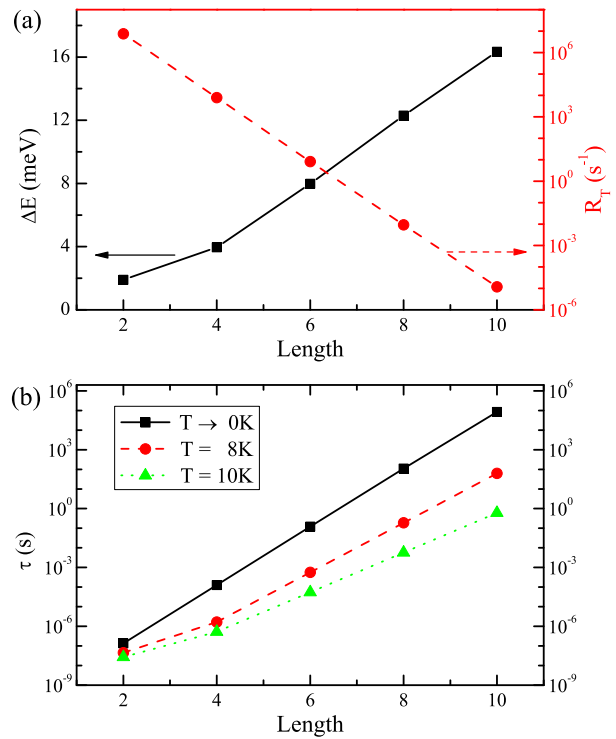}
\caption{The energy barrier ($\Delta E$) and the residual switching
rates ($R_T$) (a), and the life times ($\tau$) (b) of the $1\times N$
Fe linear chains, with $N$ taking even numbers. } \label{fig:3}
\end{figure}

\section{Manipulation of N\'{e}el states}

\begin{figure}[tbp]
\includegraphics[width=8cm]{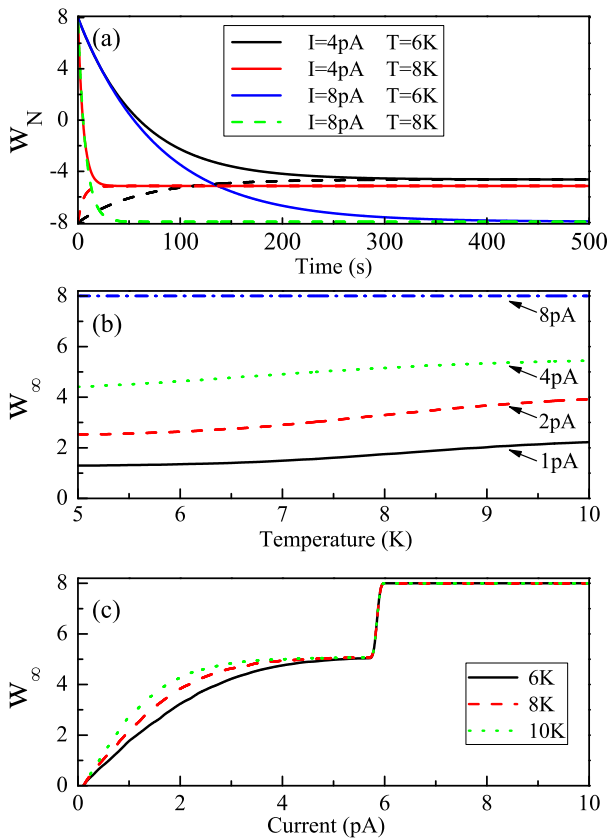}
\caption{Time dependence of the N\'{e}el weights ($W_N$) for the
$1\times 8$ chain (a), temperature dependence for different current
intensity (1, 2, 4, and 8 pA) (b) and current dependence for
different temperature (6, 8, 10 K) (c) of the target N\'{e}el
weights ($W_{\infty}$) in the same chain.} \label{fig:4}
\end{figure}

A current injected through
the STM tip on the first adatom spin can change the N\'{e}el states in the Fe spin chains\cite{AFM4,THEORY5}. To
characterize a N\'{e}el state, we define a N\'{e}el weight $W_N$ of
the $1\times N$ chain to be $\sum_i(-1)^{i-1}S_i/S$ with $i$ running
from 1 to $N$. It is $N$ and $-N$ for the N\'{e}el states
$\acute{E}_{0}$ and $\acute{E}_{1}$, respectively. We present the
N\'{e}el weight $W_N$ of $1\times 8$ spin chain under different
current intensities and temperatures in Fig. \ref{fig:4}(a). It can
be seen that the N\'{e}el weight changes exponentially with time.
The temperature and the spin-polarized current play the key role in
determining the switching rates and target N\'{e}el weights, but the
initial status of N\'{e}el weights, $\eta_0$, do not affect the
target N\'{e}el weights ($W_{\infty}$). Generally, we can fit the
time-dependent N\'{e}el weights with a simple function:
\begin{equation}
 W_N=W_0\exp (-t/t_{0})+W_{\infty}.
\end{equation}
The $t_0$ parameter describes the time scale, and $W_0+W_{\infty}$
is equivalent to the initial N\'{e}el weight. The calculated three
fitting parameters are compared in Table ~(\ref{tab:2}) between
different temperatures, different currents, and different initial
N\'{e}el weights. The temperature dependence of the target N\'{e}el
weights for four currents (1, 2, 4, and 8 pA) is presented in Fig.
\ref{fig:4}(b). The current dependence of the target N\'{e}el
weights for different temperatures (6, 8, and 10 K) is presented in
Fig. \ref{fig:4}(c). It is clear that the temperature dependence is weak,
but the target N\'{e}el weight increases substantially when the
current intensity increases. It is interesting that there exists a
big $W_{\infty}$ step at $\sim$6pA. It is because one more spin
channel becomes available at the equivalent voltage for inelastic
tunneling of the electrons\cite{ETA4,add28}.

\begin{table}
\caption{The different fitting parameters for different currents,
temperatures, and initial status. } \label{tab:2}
\begin{ruledtabular}
\begin{tabular}{cc|ccccc}
&   & $T$(K) & $W_0$(s$^{-1}$) & $t_{0}$(s) & $W_{\infty}$(s$^{-1}$)  &\\
\hline
&   $I$=4pA, $\eta_{0}$=4 & 6    & $12.24$   & $59.77$    & $-4.24$ & \\
&    & 8     & $12.86$   & $4.01$     & $-4.85$ & \\
\hline
& $I$=8pA, $\eta_{0}$=4  & 6     & $15.98$   & $82.35$    & $-7.96$ & \\
&    & 8     & $15.95$   & $6.57$     & $-7.95$ & \\
\hline
&  $I$=4pA, $\eta_{0}$=-4  &  6     & $-3.74$   & $59.78$    & $-4.25$ & \\
&    & 8    & $-3.16$   & $4.01$     & $-4.85$ & \\
\end{tabular}
\end{ruledtabular}
\end{table}

\begin{figure}[!tbp]
\includegraphics[width=8cm]{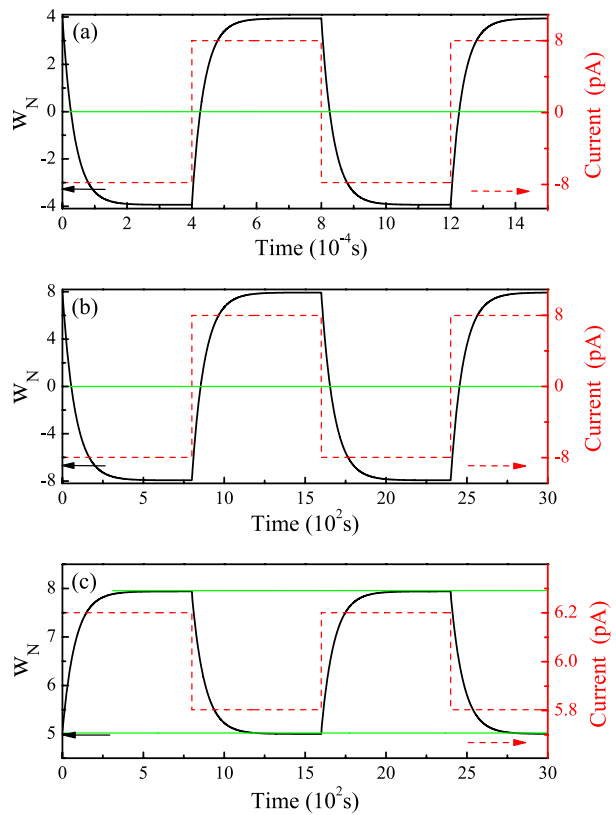}
\caption{Current manipulation of N\'{e}el states (parameterized with
$W_N$) under T = 6 K for the $1\times N$ chains with $N$=4 (a) and
$N$=8 (b,c). The current is switched between 8 pA and -8 pA for (a)
and (b), and between 5.8 and 6.2 pA for (c).} \label{fig:5}
\end{figure}

The sensitive current dependence of the target N\'{e}el weights can
be used to manipulate the N\'{e}el states through suitable
spin-polarized currents and temperatures. When a periodic
spin-polarized current between -8 and 8 pA is applied, the N\'{e}el
weights can assume a cyclic periodic function. Such manipulation of
the $1\times N$ chains ($N$ = 4, 8) with two currents with different
time periods is presented in Fig. \ref{fig:5} (a) and (b). The
N\'{e}el states are  almost fully switched for these two cases. The
two curves look similar although the time scales are hugely
different from each other. Furthermore, the big step in the
$I$-$W_{\infty}$ curves in Fig. 4(c) can be used to manipulate the
N\'{e}el states. When the current is periodically switched between
6.2 pA and 5.8 pA, the N\'{e}el weight is periodically switched
between 5.1 and 7.9, as shown in see Fig. \ref{fig:5}(c).

\section{Discussion and Conclusion}

It was pointed out by Gauyacq {\it et al}\cite{THEORY5} that there are three mechanisms for electron-induced switching between the N\'{e}el states of Fe antiferromagnetic chains. At finite temperature, thermal effects will go into action and increase exponentially with temperature. They can be attributed to a barrier-hurdling thermal activation mechanism described by an Arrhenius law.

In summary, we have investigated spin
dynamics of the antiferromagnetic spin systems through Monte Carlo
simulations. We have found out the temperature and size laws of
switching rates of N\'{e}el states for such spin systems.
Furthermore, we have shown that the N\'{e}el states can be made
stable enough for the information storage if one or two dozens of such Fe spins
are used. We also have demonstrated promising methods for
manipulating the N\'{e}el states. We believe that these can be
useful to realize antiferromagnetic information storage.

\begin{acknowledgments}
This work is supported by Nature Science Foundation of China (Grant
No. 11174359), by Chinese Department of Science and Technology
(Grant No. 2012CB932302), and by the Strategic Priority Research
Program of the Chinese Academy of Sciences (Grant No. XDB07000000).
\end{acknowledgments}


\begin{thebibliography}{99}

\bibitem{ATM1} M. L. Baker, T. Guidi, S. Carretta, J. Ollivier, H. Mutka,
H. U. Guedel, G. A. Timco, E. J. L. McInnes, G. Amoretti, R. E. P.
Winpenny, and P. Santini,  Nature Phys. \textbf{8}, 906 (2012).

\bibitem{ATM2} C. F. Hirjibehedin, C.-Y. Lin, A. F. Otte, M. Ternes,
C. P. Lutz, B. A. Jones, and A. J. Heinrich, Science \textbf{317},
1199 (2007).

\bibitem{ATM3} A. A. Khajetoorians, B. Chilian, J. Wiebe, S. Schuwalow, F. Lechermann, and R. Wiesendanger,  Nature \textbf{467}, 1084 (2012).

\bibitem{ATM4} N. Romming, C. Hanneken, M. Menzel, J. E. Bickel, B.
Wolter, K. von Bergmann, A. Kubetzka, and R. Wiesendanger, Science
\textbf{341}, 636 (2013).

\bibitem{add1} A. Spinelli, B. Bryant, F. Delgado, J. Fern\'{a}ndez-Rossier, and A.
F. Otte, Nat. Mater. 13, 782 (2014). 

\bibitem{AFM3} A. A. Khajetoorians, J. Wiebe, B. Chilian, S. Lounis, S.
Bluegel, and R. Wiesendanger,  Nat. Phys. \textbf{8}, 497 (2012).

\bibitem{AFM4} S. Loth, S. Baumann, C. P. Lutz, D. M. Eigler, and A. J. Heinrich, Science \textbf{335}, 196
(2012).

\bibitem{STM1} B. Bryant, A. Spinelli, J. J. T. Wagenaar, M. Gerrits, and A. F. Otte, Phys. Rev. Lett. \textbf{111}, 127203 (2013).

\bibitem{STM2} A. A. Khajetoorians, B. Baxevanis, C. Huebner, T. Schlenk,
S. Krause, T. O. Wehling, S. Lounis, A. Lichtenstein, D. Pfannkuche,
J. Wiebe, and R. Wiesendanger, Science \textbf{337}, 55 (2013).

\bibitem{THEORY1} W. Echtenkamp and C. Binek, Phys. Rev. Lett. \textbf{111}, 187204 (2013).

\bibitem{THEORY2} F. Meinert, M. J. Mark, E. Kirilov, K. Lauber, P. Weinmann, A. J. Daley, and H.-C. Ngerl, Phys. Rev. Lett. \textbf{111}, 053003
(2013).

\bibitem{THEORY3} W. Echtenkamp and Ch. Binek, Phys. Rev. Lett. \textbf{111}, 187204 (2013).

\bibitem{THEORY4} M. Misiorny and J. Barnas, Phys. Rev. Lett. \textbf{111}, 046603
(2013).

\bibitem{THEORY5} J.-P. Gauyacq, S. M. Yaro, and X. Cartoix\`{a}, Phys. Rev. Lett. \textbf{110}, 087201 (2013).

\bibitem{glauber} R. J. Glauber, J. Math. Phys. (N.Y.) \textbf{4}, 294
(1963).

\bibitem{ArrheniusLaw2} R.~D. Kirby, J.~X. Shen, R.~J. Hardy, and D.~J.
Sellmyer, Phys. Rev. B \textbf{49}, 10810 (1994).

\bibitem{dmc1} H. C. Kang and W. H. Weinberg, J. Chem. Phys. \textbf{90},
2824 (1989).

\bibitem{dmc2} K. A. Fichthorn and W. H. Weinberg, J. Chem. Phys. \textbf{95}%
, 1090 (1991).

\bibitem{lbg} G.-B. Liu and B.-G. Liu, Phys. Rev. B \textbf{82}, 134410
(2010).

\bibitem{ABT1} J. W. Nicklas, A. Wadehra, and J. W. Wilkins, J. Appl. Phys. \textbf{110}, 123915 (2011).

\bibitem{ETA1} N. Lorente and J.-P. Gauyacq, Phys. Rev. Lett. \textbf{103}, 176601 (2009).

\bibitem{ETA2} J. Fern\'{a}ndez-Rossier, Phys. Rev. Lett. \textbf{102}, 256802 (2009).

\bibitem{ETA3} A. X. S\'{a}nchez and J.-P. Leburton, Phys. Rev. B \textbf{88}, 075305 (2013).

\bibitem{ETA4} J. P. Gauyacq and N. Lorente, Phys. Rev. B \textbf{87}, 195402 (2013).

\bibitem{add28} F. D. Novaes, N. Lorente, and J.-P. Gauyacq, Phys. Rev. B \textbf{82}, 155401
(2010). 

\end{thebibliography}
\end{document}